\documentclass[12pt,preprint]{aastex}
\usepackage{txfonts}
\usepackage{subfigure}
\usepackage{multirow}

\usepackage{times}
\usepackage{natbib}
\usepackage{lscape}
\usepackage{rotating}
\usepackage{url}
\usepackage{amssymb}
\bibpunct{(}{)}{;}{a}{}{,}

\begin{document}

\title{The link between the Hidden Broad Line Region and the accretion rate in Seyfert 2 galaxies}

\author{Andrea Marinucci\altaffilmark{1,2}, Stefano Bianchi\altaffilmark{1,3,4}, Fabrizio Nicastro\altaffilmark{2,4,5}, Giorgio Matt\altaffilmark{1}, Andy D. Goulding\altaffilmark{2}}


\altaffiltext{1}{Dipartimento di Fisica, Universit\`a degli Studi Roma Tre, via della Vasca Navale 84, 00146 Roma, Italy}
 \altaffiltext{2}{Harvard-Smithsonian Center for Astrophysics, 60 Garden St., Cambridge MA 02138, USA}
 \altaffiltext{3}{INAF - Osservatorio Astronomico di Brera, Via E. Bianchi 46, I-23807, Merate, Italy}
\altaffiltext{4}{IESL, Foundation for Research and Technology, 711 10, Heraklion, Crete, Greece}
 \altaffiltext{5}{Osservatorio Astronomico di Roma (INAF), Via Frascati 33, I-00040 Monte Porzio Catone, Italy}

\date{Received / Accepted}

\begin{abstract} 
In the past few years more and more pieces of evidence have been presented for a revision of the widely accepted Unified Model of Active Galactic Nuclei. A model based solely on orientation cannot explain all the observed phenomenology. 
In the following, we will present evidence that accretion rate is also a key parameter for the presence of Hidden Broad Line Regions in Seyfert 2 galaxies.
Our sample consists of 21 sources with polarized Hidden Broad Lines and 18 sources without Hidden Broad Lines. 
We use stellar velocity dispersions from several studies on the Ca\textsc{ii} and Mg b triplets in Seyfert 2 galaxies, to estimate the mass 
of the central black holes via the $M_{bh}$-$\sigma_{\star}$ relation. 
The ratio between the bolometric luminosity, derived from the intrinsic (i.e. unabsorbed) X-ray luminosity, and the Eddington luminosity is 
a measure of the rate at which matter accretes onto the central supermassive black hole. 
A separation between Compton-thin HBLR and non-HBLR sources is clear, both in accretion rate ($\log L_\mathrm{bol}/L_\mathrm{Edd}=-1.9$)  
and in luminosity ($\log L_\mathrm{bol}=43.90$).  
When, properly luminosity-corrected, Compton-thick sources are included, the separation between HBLR and non-HBLR is less 
sharp but no HBLR source falls below the Eddington ratio threshold. We speculate that non-HBLR Compton-thick sources with 
accretion rate higher than the threshold, do possess a BLR, but something, probably related to their heavy absorption, is preventing 
us from observing it even in polarized light. 

Our results for Compton-thin sources support theoretical expectations. In a model presented by Nicastro (2000), the presence of broad 
emission lines is intrinsically connected with disk instabilities occuring in proximity of a transition radius, which is a function of the accretion rate, 
becoming smaller than the innermost stable orbit for very low accretion rates and therefore luminosities. 

\end{abstract}


\section{Introduction}
The widely accepted unification model for active galactic nuclei (AGNs) invokes the same paradigm for Seyfert 2 and Seyfert 1 galaxies 
\citep{antonucci93}. The two different types of galaxies are believed to be intrinsically the same but differ, observationally, due only to 
orientation: Seyfert 1 galaxies are observed at large accretion disk-line of sight angles (face-on) while Seyfert 2 galaxies are seen edge-on 
through large column densities of obscuring material in the accretion disk plane, which prevent the direct view of the nuclear regions 
of these sources. 
This scenario came into existence after the discovery of polarized broad permitted emission lines in one of the brightest Seyfert 2 galaxies, 
NGC 1068 \citep{antonucci85, miller83}, suggesting a geometry in which: (a) BLRs are confined in a relatively small region ($\sim$ light-days) 
surrounding the central source, (b) their direct view (in Seyfert 2s) is obscured by a flat distribution of distant material coplanar with the 
disk-plane, and (c) their line emission is Compton-scattered into the line of sight direction off a population of hot electrons extending 
at large radii above and below the accretion disk.
 
In the past few years, exceptions to orientation-based unification models have been found, suggesting the possibilty that not all Seyfert 2 galaxies host a Seyfert 1 nucleus. Indeed, spectropolarimetric surveys find that only about the 50$\%$ of the brightest Seyfert 2 galaxies show hidden broad-line regions (HBLR) in their optical-polarimetric spectra \citep{tran01,tran03}. 
Several authors have suggested that the presence of BLRs in Seyfert 2s can be linked to the luminosity of the active nucleus, and may 
disappear at low luminosities  \citep{lumsden01,tran01,guhuang02,mama02,tran03,laor03,eliho09} or low accretion rates 
\citep{nic00,nmm03,cze04} .  
In both cases, the presence of the BLR is not an inescapable feature of all Seyfert galaxies, as postulated by the unification model, but it is 
tightly linked to a physical parameter of the AGN, either the luminosity or the accretion rate. 
 However, recent works, casted doubts on these conclusions \citep[e.g.][]{biangu07}. In these works large, but inhomogeneous, samples were 
used to identify a physical parameter responsible for the existence or absence of HBLR in AGNs, but no clear indication for the existence of 
such a parameter was found. We think that this is largely due to the use of the [OIII] luminosities as a proxy of the nuclear activity of the 
AGN, and the difficulty in correcting this for extinction in obscured objects. 
It has been shown by several author, by comparing AGN luminosities derived from [OIII] and [OIV] emission line luminosities, that the 
observed [OIII] luminosities often suffer significant attenuation in typical Seyfert 2 galaxies \citep[e.g.][]{haas05, melendez08,diamond09, goual09, baum10, krae11} and that the efforts to use standard 
extinction correction for [OIII] are not always reliable \citep[e.g.][]{goual09, lamassa10}. 

[OIV] emission line luminosities might therefore be more reliable than [OIII] luminosities, as tracers of the intrinsic nuclear emission. However, 
they are still an indirect proxy of the nuclear continuum emission. 
In this paper, instead, we use the observed X-ray (2-10 keV) continuum emission as a direct probe of the AGN activity. We 
selected a sample of type 2 AGN with good quality spectropolarimetric and X-ray observations, for which we can give a good estimate 
of the mass of the central supermassive black hole (BH).  \footnote{Our work is based on a previous study discussed in \citet{nmm03}, but here we use a wider sample and more robust estimates of the BH masses.}
By doing so, we find evidence suggesting that accretion rate is the main parameter that sets the existence of HBLRs in Seyfert 2 
galaxies. 

\section{The Sample}
Our starting sample is mostly based on the spectropolarimetric surveys performed by \citet{tran95,tran01,tran03} on the Seyfert 2 
galaxies included in the CfA \citep{hb92} and 12 $\mu$m \citep{rms93} samples. 
Additionally, we included objects from other, high-quality, spectropolarimetric studies \citep{tran92, young96, moran00, lumsden01}. 
From this sample, consisting of 90 candidates, we selected only sources with available mass (Section 2.1) and bolometric luminosity 
(Section 2.2) estimates, and so for which their accretion rate can be evaluated. 
For each source, we evaluate their accretion rate in units of Eddington, as the ratio $L_{bol}/L_{Edd}$ (hereafter $\lambda_{Edd}$), where
\begin{equation}
L_{Edd}=1.2\times10^{38} \left( \frac{M_{BH}}{M_\odot} \right)\, \mathrm{erg\,s^{-1}} .
\end{equation}

\begin{sidewaystable}
\caption{\label{samplelist}\footnotesize The sample used in this paper. X-ray and bolometric luminosities are already corrected for a factor 70 for all Compton Thick sources. Columns: (a) Name of the object;
(b) Instrument (AS: \textit{Chandra} ACIS-S; PN: XMM-\textit{Newton} EPIC pn; XIS: \textit{Suzaku} XIS0+3) and \textsc{obsid};
(c) \checkmark if Compton-thin, X if Compton-thick;
(d) Stellar velocity dispersions;
(e) Mass of the central BH in solar mass units;
(f) Absorbing column density in $10^{22}$ cm$^{-2}$ units
(g) Absorption corrected 2-10 keV luminosity;
(h) Bolometric luminosity;
(i) Accretion rate in Eddington units;
(l) Extinction-corrected O [\textsc{iii}] ($\lambda\ 5007$ ) luminosity;
(m) O [\textsc{iv}] ($\lambda25.89\ \mu m$ ) luminosity
(n) Reference for stellar velocity dispersions, HBLR (in order); 
{\bfseries References}: (1) \citet{gr05}, (2) \citet{nw95}, (3) \citet{tdt90}, (4) \citet{oomm99}, (5) \citet{bhs02}, (6) \citet{swc93}, (7) \citet{mce95}, (8) \citet{tran03}, (9 )\citet{tran95}, (10) \citet{moran00}, (11) \citet{young96}, (12) \citet{lumsden01}, (13) Other surveys}

\begin{center}
\scriptsize
Hidden Broad Line Region Seyfert 2s\\
\begin{tabular}{cccccccccccc}
{\bfseries Object name } & {\bfseries Obs. ID }& {\bfseries C-Thin } & $\sigma_*$ (km s$^{-1}$)&{\bfseries log$(M_{bh}/M_{\odot})$}& $N_H$ &{\bfseries log$(L_{2-10\ keV})$ }&{\bfseries log$(L_{Bol})$}&{\bfseries log$(\lambda_{edd})$} & {\bfseries log$(L_{[OIII]})$ }&{\bfseries log$(L_{[OIV]})$ }&{\bfseries References }\\
(a) &(b) &(c) &(d) &(e) &(f) &(g) &(h) &(i)&(l)&(m)&(n)\\
 \hline \\
CIRCINUS &PN- 0111240101&X&$75\pm5$ & 6.42& $430_{-30}^{+40}$&42.62 &43.76&-0.75& 40.92&40.58&1, 13\\
IC 3639 & XIS-702011010&X&$99 \pm5$& 6.90& $>$150&42.64 &43.80&-1.20 &41.89&40.66&1,8\\
IC 5063 & XIS-704010010&\checkmark& $160 \pm 25$&7.74& $25^{+3}_{-2}$&42.83&44.03 &-1.81&41.56&41.40&2,8  \\
IRAS01475-0740 &PN-0200431101 &X&$108\pm17$ & 7.05&$>200$ &43.52 &  44.90&-0.25&41.76&40.74&1,8\\
MCG -2-8-39  & PN-0301150201&\checkmark & $126\pm11 $&7.32& $44^{+23}_{-24}$&42.57&43.70&-1.72 & 41.16&41.07&1,8 \\
Mrk 3 &XIS-100040010 &X&$249\pm4$ & 8.51 &$140^{+8}_{-12}$ &44.44&46.10&-0.51 &43.27&41.97&1, 10\\
Mrk 348 & PN-0067540201&\checkmark &$117 \pm 18$& 7.19& $15^{+2}_{-1}$&43.41& 44.77&-0.52&41.96&40.95&2,8\\
Mrk 1210  & PN-0002940701&\checkmark& $82\pm16 $& 6.57 &$22^{+5}_{-6}$ &43.02&44.24 &-0.43 & 42.37&-&1,9\\
NGC 513 & PN-0301150401&\checkmark& $150 \pm 25$&7.63& $7^{+2}_{-2}$&42.72&  43.93&-1.80&41.14 &40.74&2,8 \\
NGC 591 & PN-0200431001&X&$107\pm18$ & 7.04 & $>160$&43.02 &44.27&-0.87&41.97&-&2,10\\
NGC 788 & XIS-703032010&\checkmark&$140 \pm 20$& 7.51&$85^{+7}_{-6}$ &43.15&44.43&-1.18&40.73&40.97&2,10\\
NGC 1068 & PN-0111200201&X&$147\pm3$ & 7.59&$>1000$ &43.02 &44.27&-1.42& 42.38&41.81&1,8\\
NGC 2273& XIS-702003010&X&$136\pm22$ & 7.46& $120^{+110}_{-46}$&42.73 &  43.90&-1.65&41.13&40.07&2, 10\\
NGC 3081  & XIS-70301301& \checkmark& $113 \pm 4$&7.13&$89^{+13}_{-15}$ & 42.50& 43.61&-1.62 &41.43&41.09&1,10\\
NGC 4388& XIS-800017010 &\checkmark&  119& 7.22& $31^{+2}_{-2}$&42.90&44.12 &-1.20&41.85&41.58&3,8\\
NGC 4507 & XIS-702048010& \checkmark& $152 \pm 4$ &7.65& $87^{+7}_{-8}$ &43.11 &44.39 &-1.36&42.19&41.02&1,10\\
NGC 5252& PN-0152940101&\checkmark& $190 \pm 27$&8.04&$2.2^{+0.1}_{-0.1}$ &43.04& 44.30&-1.84& 42.05&-&2,11\\
NGC 5506& PN-0554170101&\checkmark& 180& 7.95&$3.0^{+0.1}_{-0.1}$  &43.05&44.30&-1.75& 41.45&41.28&4,8\\
NGC 7212 & PN-0200430201&X&$140 \pm9$& 7.51 & $>150$&43.77 &45.22&-0.38&42.73&-&1,13\\
NGC 7674 &PN-0200660101 &X &$144\pm 32$& 7.56 & $>100$&43.96 &45.47&-0.18&42.57&41.97&2,8\\
NGC 7682 & PN-0301150501&X&$123\pm17$ & 7.28 & $>100$&43.02 &44.27&-1.11&41.76&41.01&2,8\\
\hline
\end{tabular}\\
 \scriptsize
$\ \ \ \ \ $ \\
Non Hidden Broad Line Region Seyfert 2s\\
\begin{tabular}{cccccccccccc}
{\bfseries Object name } &  {\bfseries Obs. ID }&{\bfseries C-Thin} & $\sigma_*$ (km s$^{-1}$)&{\bfseries log$(M_{bh}/M_{\odot})$}& $N_H$&{\bfseries log$(L_{2-10\ keV})$ }& {\bfseries log$(L_{Bol})$}&{\bfseries log$(\lambda_{edd})$} &{\bfseries log$(L_{[OIII]})$ }&{\bfseries log$(L_{[OIV]})$ }&{\bfseries References }\\
(a) &(b) &(c) &(d) &(e) &(f) &(g) &(h) &(i)&(l)&(m)&(n)\\
 \hline
M51 & PN-0303420101& X&  $82 \pm 11$&6.57 & $>400$ &41.54& 42.51&-2.17& 40.03&39.01&2,8\\
Mrk 573& AS-7745&X& $148 \pm 3$&7.60&$>100$ &43.10 &  44.37&-1.33& 42.39&41.71&1,8\\
Mrk 1066& PN-0201770201&X&$119 \pm 19$&7.22&$>100$ &42.92&44.15&-1.17& 42.27&-&2,10\\
NGC 1320& PN-0405240201& X&$124 \pm 14$& 7.29& $>100$&42.69&43.86&-1.54&41.08 &40.67&2,8\\
NGC 1358& PN-0301650201&X &$185 \pm 20$ &7.99&$130^{+850}_{-60}$ &43.05& 44.31&-1.79& 41.36&40.56&2,10\\
NGC 1386& XIS-702002010&X& $133 \pm 3$&7.42 &$>100$ &41.62& 42.59& -2.92& 41.09&40.25&1,8\\
NGC 1667& XIS-701006010&X&173&7.88&$>100$  &42.56&43.70&-2.28& 42.03&40.51&3,8\\
NGC 3079& XIS-803039020&X& $150 \pm 10$&7.63& $>100$ &42.02&43.05&-2.68& 40.48&39.53&6,8\\
NCG 3281 & XIS-703033010&\checkmark & $176 \pm 3$ & 7.91 & $79^{+10}_{-7}$&42.65&43.81&-2.20& 41.30&41.58&1,10\\
NGC 3982& PN-0204651201 &X &$81 \pm 13$&6.55& $>100$&41.15& 42.06&-2.59&40.33 &39.10&5,8\\
NGC 5135& XIS-702005010&X&$124 \pm 6$&7.29& $>100$ &43.10& 44.37&-1.02& 42.21&41.48&1,8\\
$\ \ \ \ \ \ $NGC 5283 $\ \ \ \ \ $& AS-4846&\checkmark & $148 \pm 14$&7.60 &$12^{+3}_{-3}$ &41.54& 42.49&-3.21&40.88 &-&2,8\\
NGC 5347& XIS-703011010&X&$103 \pm 14$&6.97&$>100$ &42.39&  43.49&-1.58& 41.22&40.01&3,8\\
NGC 5728& XIS-701079010 &X &209&8.21& $>150$&43.29&44.61 &-1.69& 42.83&41.36&7,11\\
NGC 7582 & PN-0112310201&\checkmark &  $113 \pm 3$& 7.13 & $57^{+10}_{-10}$& 41.90&42.91 &-2.32& 41.63&41.11&1,8\\
NGC 7130& XIS-703012010&X &$147 \pm 5$&7.59& $>100$&43.10& 44.37&-1.32& 42.55&40.94&1,12\\
NGC 7172 & PN-0202860101&\checkmark & $160 \pm 9$&7.74 & $8.4^{+0.1}_{-0.1}$ &42.73& 43.90&-1.93& 40.84&40.82&1,8\\
UGC 6100& PN-0301151101&X&$156 \pm 25$& 7.69& $>100$&42.84&44.05&-1.75&42.18 &-&2,8\\
\hline
\end{tabular}
\end{center}

\end{sidewaystable}
\subsection{The BH mass sub-selection}
In this work, BH masses are homogenuously derived for the entire sample, by using uniquely the $M_{BH}$-$\sigma_{\star}$ relation 
\citep{trem02}:

\begin{equation}
M_{BH}=1.35\times 10^8\ \Big(\frac{\sigma_*}{200\ km\ s^{-1}} \Big)^{4.02}\ M_{\odot}.
\end{equation}

Stellar velocity dispersions ($\sigma_*$) were mainly taken from \citet{nw95} and \citet{gr05}. 
These estimates are all based on direct meaurements of Ca\textsc{ii} (8498 $\r{A}$, 8542 $\r{A}$ and 8662 $\r{A}$) and 
Mg\textsc{i} $3p\rightarrow 4s$ (5167 $\r{A}$, 5172 $\r{A}$ and 5183 $\r{A}$, hereinafter Mg {\itshape b}) triplet absorption, 
imprinted by the interstellar medium of Seyfert 2 galaxies. 
For sources with more than one measurement, we used that with smaller error bars. 
Uncertainties on the BH mass estimates based on the $M_{BH}$-$\sigma_{\star}$ relation, come from the statistical errors on the 
$\sigma_{\star}$ measurements (listed in Table 1) as well as from the spread in the $M_{BH}$-$\sigma_{\star}$ relation itself,  estimated in 
0.44 dex \citep{gulte09}. This spread is generally much larger than the statistical error on $\sigma_{\star}$. 
We therefore assumed an uncertainty of 0.44 dex for all our BH mass estimates. 
Stellar velocity dispersions and their associated uncertainties, as well as BH masses, 
are listed in Table~\ref{samplelist} (columns (d) and (e), respectively). 

The $\sigma_*$ selection reduced the original spectropolarimetric sample to 46 sources with direct $\sigma_*$ measurements. 

\subsection{\label{xanalysis} The X-Ray Sub-Selection and the Final Sample}
To estimate the bolometric luminosity of the sources of our sample, we use the 2-10 keV luminosity, which is a direct tracer of the 
primary emission, and apply a bolometric correction. 
We searched  the XMM-\textit{Newton}, \textit{Chandra}, \textit{Suzaku} (the front-illuminated CCDs have the largest area at high 
energies), and \textit{Swift} archives for observations of the sources of our sample of 46 objects with accurate BH-mass estimate. 
When multiple observations of a single target were available, we selected that with the highest S/N ratio in the 2-10 keV band. 
For all sources of our final sample, a minimum of 150 counts in the 2-10 keV band was required, to derive reliable estimates of 
the column density and, therefore, intrinsic X-ray luminosity. 
These criteria reduced the final sample to a total of 39 sources: 21 with polarized hidden Broad Emission Lines (HBLRs) and 
18 sources without (non-HBLRs). These are listed in Table~\ref{samplelist}, together with 
the selected X-ray observations. 

\section{Data Analysis}

\subsection{X-ray Data Reduction and Analysis} 
\textit{Suzaku} X-ray Imaging Spectrometer (XIS) data were processed with the latest calibration files available 
at the time of the analysis (2011-02-10 release), by using FTOOLS 16 and SUZAKU software Version 2.3, and adopting standard 
filtering procedures. Response matrices and ancillary response
files were generated using \textsc{XISRMFGEN} and \textsc{XISSIMARFGEN}. 
The 0.5-10 keV spectra extracted from the front-illuminated XIS0 and XIS3 have been co-added via the ftool \textsc{addascaspec}.

We used only XMM-\textit{Newton} observations performed with the EPIC-Pn camera \citep{struder01} operated in large window and 
medium filter modes. Source data 'cleaning' (exclusion of flaring particle background intervals) and spectra extraction, were performed 
with SAS 10.0.0 \citep{gabr04} via an iterative process which leads to a maximization of the Signal-to-Noise Ratio (SNR), similarly to 
that described in \citet{pico04}. For each source, background spectra were extracted from source-free circular regions of the source field. 

Finally, \textit{Chandra} data were reduced with the Chandra Interactive Analysis of Observations \citep[CIAO;][]{ciao} 4.3 and 
the Chandra Calibration Data Base (CALDB) 4.4.1 software, by adopting standard procedures.

All the spectra with a high S/N ratio were binned in order to over-sample the instrumental resolution by at least a factor of 3 and to 
have no less than 30 counts in each background-subtracted spectral channel. This allows the applicability of the $\chi^2$ statistics. 
The adopted cosmological parameters are $H_0=70$ km s$^{-1}$ Mpc$^{-1}$, $\Omega_\Lambda=0.73$ and $\Omega_m=0.27$. 
Errors are quoted at a confidence level of 90\% for one interesting parameter ($\Delta\chi^2=2.7$), if not otherwise stated. 
The spectral analysis was performed with the package \textsc{xspec 12.7.0} \citep[][]{xspec}. 

For all sources of our sample, we fit their 2-10 keV spectra with a general baseline model, consisting of a power law continnuum 
atennuated by the line of sight column of Galactic absorbtpion plus intrinsic absorption at the source redshift, plus three additional 
emission components: (a) a photoionized plasma emitter, to model the soft excess often detected in Seyfert 2s at E$\lesssim 2$ keV, 
(b) a cold Compton Reflector, scattering the primary nuclear photons off the inner walls of cold circum-nuclear material and along 
the line of sight, and (c) positive gaussian profiles, to model fluorescence emission lines of high-Z elements such as Fe at 6.4 keV, as 
required by the data. 
The model can be parameterized as: 

\begin{equation}
 F(E)= e^{-\sigma(E) N_H^G}[Ph_C + e^{-\sigma(E) N_H}BE^{-\Gamma} + R(\Gamma)+ \sum_i G_i(E)]
\end{equation}
where $\sigma(E)$ is the photoelectric cross-section \citep[abundances as in][]{angr89}, $N_H^G$ is the line-of-sight Galactic 
column density \citep{dl90}; $Ph_C$ is the photoionized plasma emission \citep[see][for details on the adopted \textsc{cloudy} 
model]{bianchi10}; $N_H$ is the neutral absorbing column density at the redshift of the source; 
B is the normalization of the primary powerlaw with slope $\Gamma$; $R(\Gamma)$ is the Compton-reflection component 
(modelled in \textsc{Xspec} with {\sc pexrav} \citep{mz95}); and $G_i(E)$ are the required Gaussian profiles. 
\begin{figure*}[h]
\centering
\includegraphics[width=5in]{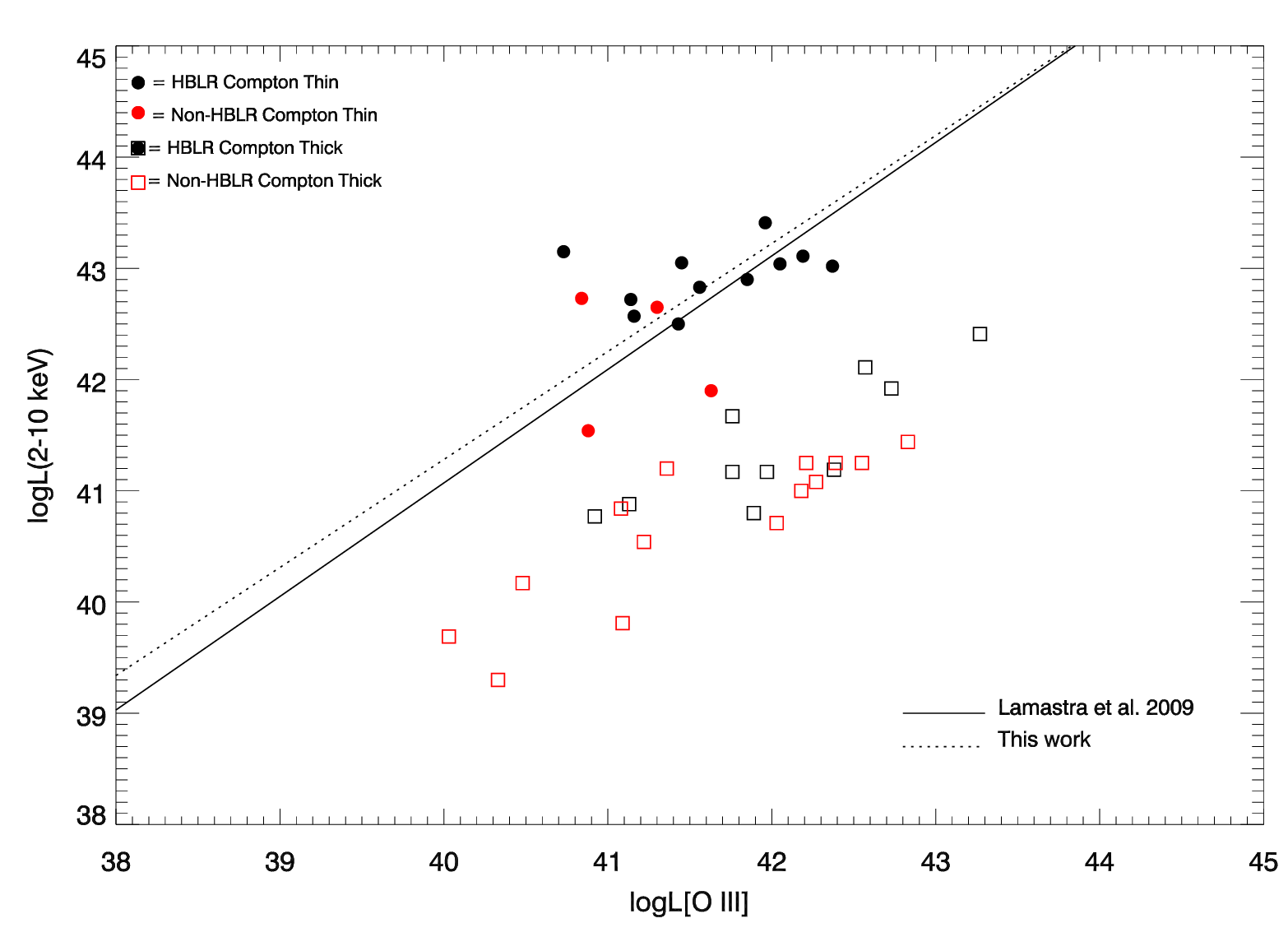}
\includegraphics[width=5in]{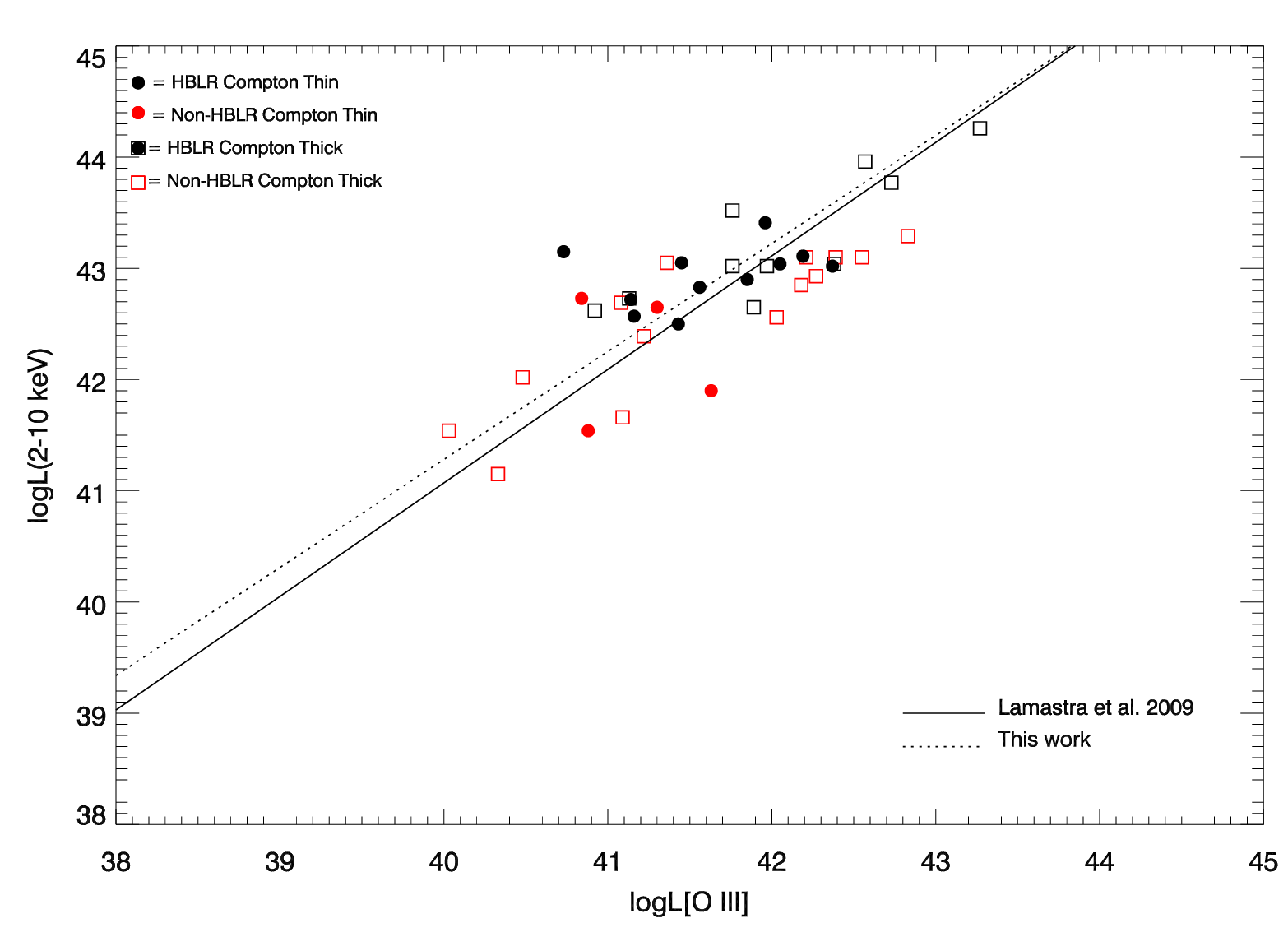}
\caption{\label{lumx_lumo} 2-10 keV luminosities derived from our x-ray analysis are plot against [O\textsc{iii}] luminosities. A correction factor of 70 is used in the bottom panel.}
\end{figure*}

Our adopted models do not aim at obtaining the `best fit' for each source, but only a reliable estimate of the observed 
2-10 keV luminosity (Table~\ref{samplelist}, column (g)). 
A more detailed description of the results of the fits, in particular for the photoionization \textsc{cloudy} component, is deferred to a 
future work.

\subsection{Nuclear 2-10 keV Luminosity for Compton-Thick Sources}
Compton-Thick sources are defined as sources for which the intrinsic column of absorbing gas, $N_H$, is greater than 
$\simeq 10^{24}$ cm${}^{-2}$, so preventing us from directly observing the primary powerlaw below 10 keV. 
The 2-10 keV spectra of Compton-thick sources are therefore generally adequately modeled by a pure-reflection component (i.e. reflection-dominated Seyfert 2s), which can only provide a measurement of the reflected, and not the nuclear, luminosity. \\
 Several tracers can be used to estimate the bolometric luminosity of Compton-thick objects, e.g. [O\textsc{iii}] \citep{lamastra09} and [O\textsc{iv}] \citep{rigby09} . For homogeneity with the Compton-thin sources of our sample, here we use the observed X-ray luminosity, by applying the following procedure.

 We compare the extinction-corrected [O\textsc{iii}] luminosities of the sources of our sample (Table~\ref{samplelist}, 
column (l)), taken from \citet{wu11} and \citet{biangu07}, with their observed 2-10 keV reflection-dominated luminosities. 
Both works correct the observed $F_{[OIII]}$ for extinction, by applying the relation in \citet{bass99} :
\begin{equation}
F_{[O III]}^{cor}=F_{[OIII]}^{obs}\Big{[}\frac{(H\alpha/H\beta)_{obs}}{(H\alpha/H\beta)_{0}}\Big{]}^{2.94},
\end{equation} 
where an intrinsic Balmer decrement $(H\alpha/H\beta)_{0}$=3 is adopted.
The results are shown in Fig.~\ref{lumx_lumo}, top panel. 
All the Compton-Thick sources have a $L_x/L_{[O\textsc{iii}]}$ ratio smaller than the best fitting relation found by \citet{lamastra09} 
for a large sample of Compton-thin Seyfert 2s (solid line in Fig.~\ref{lumx_lumo}: compare with the dotted line in the same 
figure, which best-fits the sub-sample of Compton-thin sources of our sample). 
For the Compton-thick sources of our sample, we find a mean value of $<\log(L_X/L_{[O\textsc{iii}]})>=-0.76\pm0.09$. 
For the Compton-thin sources of their sample, instead, \citet{lamastra09} found a mean $<\log(L_X/L_{[O\textsc{iii}]})>=1.09$. 
The ratio between these two means can be used as an estimate of the correction factor needed to infer the nuclear 2-10 keV 
luminosity of Compton-thick sources, from their observed 2-10 keV luminosity: $10^{1.09-(-0.76)}\simeq70$. 
The bottom panel of Fig.~\ref{lumx_lumo} shows the $logL_X$ - $logL_{[O\textsc{iii}]}$ plane for all the sources of our sample, after applying 
a correction factor of 70 to the observed 2-10 keV luminosity of all Compton-thick sources. 
Now all the sources of our sample lie on the \citet{lamastra09} relation for Compton-thin sources.

We decided to adopt this correction for all Compton-thick sources, also for those where E$>$10 keV data were 
available (\textit{Suzaku} PIN) and so a direct estimate of the nuclear continuum above 10 keV could in principle be attempted. 
This is because Compton scattering at high column densities may significantly suppress the observed intrinsic luminosity even 
at E$>$ 10 keV, and the modellization of these effects are highly dependent on the (unknown) geometry of the absorber \citep[e.g.][]{mpl99}. 
In the cases where objects presented observations in both Compton-thin and Compton-thick states in the past few years we chose the 
former, to infer with a higher precision the unabsorbed luminosity of the source.
In one case (IRAS01475-0740 ), the 2-10 keV source spectrum is ambiguous: following the detailed analysis presented in \citet{brna08} 
we classify this object as Compton-thick.
  
\subsection{Bolometric Luminosities and Eddington Ratios}
For each source of our sample, we used their intrinsic 2-10 keV luminosity (observed, for Compton-Thin sources, or inferred, for 
Compton-Thick sources, see previous section) to derive their bolometric luminosity, and thus Eddington ratio (Table~\ref{samplelist}, 
columns (h) and (i)), by adopting the luminosity-dependent relation presented in \citet{mar04}. 
We tested the importance of this assumption in our calculations, by trying different bolometric corrections \citep[e.g.][]{elvis94,vf09}, 
and comparing the results. The most significant differences, between these methods, were found at the extremes of the luminosity 
(accretion) range spanned by the sources of our sample. Therefore, the exact choice of the bolometric correction does not significantly 
affect our results. 

The major contribution to the bolometric luminosity error determination comes from the uncertainty on the bolometric correction itself, 
which is generally significantly larger than the uncertainty on the X-ray luminosity. The bolometric correction proposed by \citet{mar04} is 
based on the correlation between the UV luminosity at 2500 $\r{A}$ and the 2-10 keV X-ray luminosity, whose spread is estimated 
as 0.37 dex \citet[see eq. 3 of][]{yer10}. Therefore, we assume this uncertainty on our derived bolometric luminosities.

\begin{table*}
\caption{\label{kstests}Results from the two-sample K-S test. For Compton Thick sources a correction factor of 70 is used for the calculation of the 2-10 keV luminosity and hence for the bolometric luminosity. 
(1) Description of the sample; (2) number of objects in the sample; (3) percentage probability that the two samples derive from the 
same parent population in $L_{Bol}$; (4) percentage probability that the two samples derive from the same parent population in 
$L_{2-10\ keV}$, $L_{[O III]}$, or $L_{[O IV]}$; (5) percentage probability that the two samples derive from the same parent population in 
$\lambda_{edd}$; (6),(7),(8) maximal separation value between the cumulative distributions of the two samples.}
\begin{center}
\scriptsize
\begin{tabular}{cccccccc}
{\bfseries Sample } & N & P($L_{Bol}$)  &  P($L_{2-10\ keV}$)  &P($\lambda_{edd}$)  & $L_{Bol}^{max}$ &  $L_{2-10\ keV}^{max}$ & $\lambda_{edd}^{max}$  \\
(1) &(2) &(3) &(4) & (5) & (6) &(7) &(8) \\
\hline
 & & & \\
Non-HBLR Compton Thin& 4& \multirow{2}{*}{0.5\%}& \multirow{2}{*}{4.4\%}& \multirow{2}{*}{0.1\%}&\multirow{2}{*}{43.90}& \multirow{2}{*}{42.73}&\multirow{2}{*}{-1.9}\\
HBLR Compton Thin  &11 & & \\
 & & & \\
Non-HBLR Compton Thick and Thin& 18& \multirow{2}{*}{1.8\%}& \multirow{2}{*}{6.9\%}& \multirow{2}{*}{1.0\%}& \multirow{2}{*}{43.90}& \multirow{2}{*}{42.56} &\multirow{2}{*}{-1.8}\\
HBLR Compton Thick and Thin & 21& & \\
 & & & \\ 
{\bfseries [OIII] }& & & P($L_{[OIII]})$  & & &$L_{[OIII]}^{max}$ \\
\hline
 & & & \\
Non-HBLR Compton Thin&4& \multirow{2}{*}{38.5\%}& \multirow{2}{*}{ 38.5\%}& \multirow{2}{*}{8.7\%}&  \multirow{2}{*}{43.51}&\multirow{2}{*}{41.36} &\multirow{2}{*}{-2.5}\\
HBLR Compton Thin &11 & & \\
 & & & \\
Non-HBLR Compton Thick and Thin& 18& \multirow{2}{*}{23.2\%}& \multirow{2}{*}{ 23.2\%} & \multirow{2}{*}{25.7\%}& \multirow{2}{*}{43.54}& \multirow{2}{*}{ 41.39}&\multirow{2}{*}{-2.1}\\
HBLR Compton Thick and Thin &21 & & \\
 & & & \\
 {\bfseries [OIV] }& & & P($L_{[OIV]})$  & & &$L_{[OIV]}^{max}$ \\
\hline
 & & & \\
Non-HBLR Compton Thin&3& \multirow{2}{*}{-}& \multirow{2}{*}{ -}& \multirow{2}{*}{-}&  \multirow{2}{*}{-}&\multirow{2}{*}{-} &\multirow{2}{*}{-}\\
HBLR Compton Thin &9& & \\
 & & & \\
Non-HBLR Compton Thick and Thin& 15& \multirow{2}{*}{10.4\%}& \multirow{2}{*}{ 10.4\%} & \multirow{2}{*}{1.1\%}& \multirow{2}{*}{43.82}& \multirow{2}{*}{40.57}&\multirow{2}{*}{-1.4}\\
HBLR Compton Thick and Thin &17& & \\
 & & & \\
\hline
\end{tabular}
\end{center}
\end{table*}

For the Eddington ratios, we propagated the uncertainties on the BH mass and bolometric luminosity. This leads to an uncertainty 
of 0.5 dex in the Eddington ratios.

\section{\label{results}Results}
\begin{figure*}[t!]
\centering
\includegraphics[width=5in]{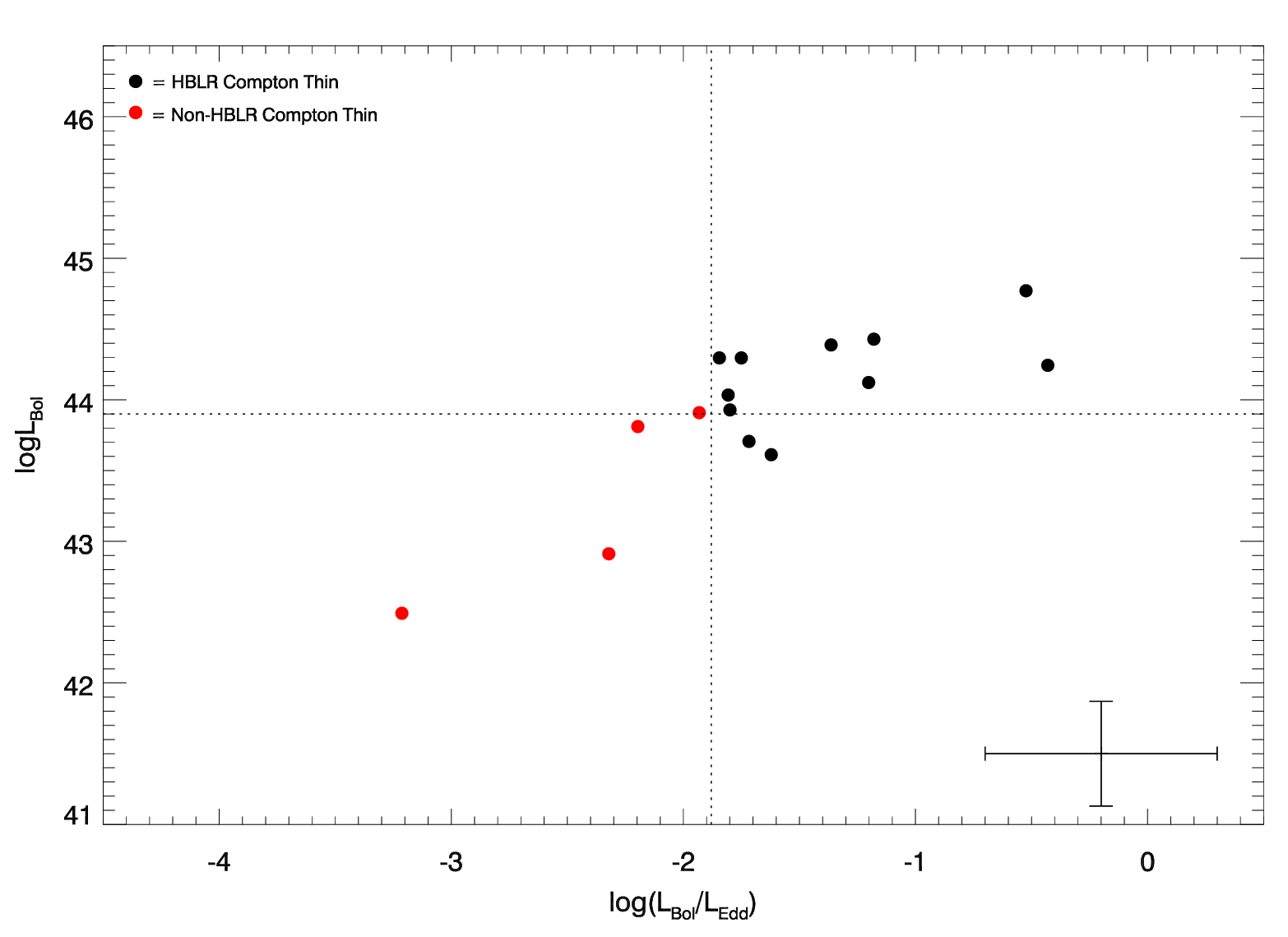}
\includegraphics[width=5in]{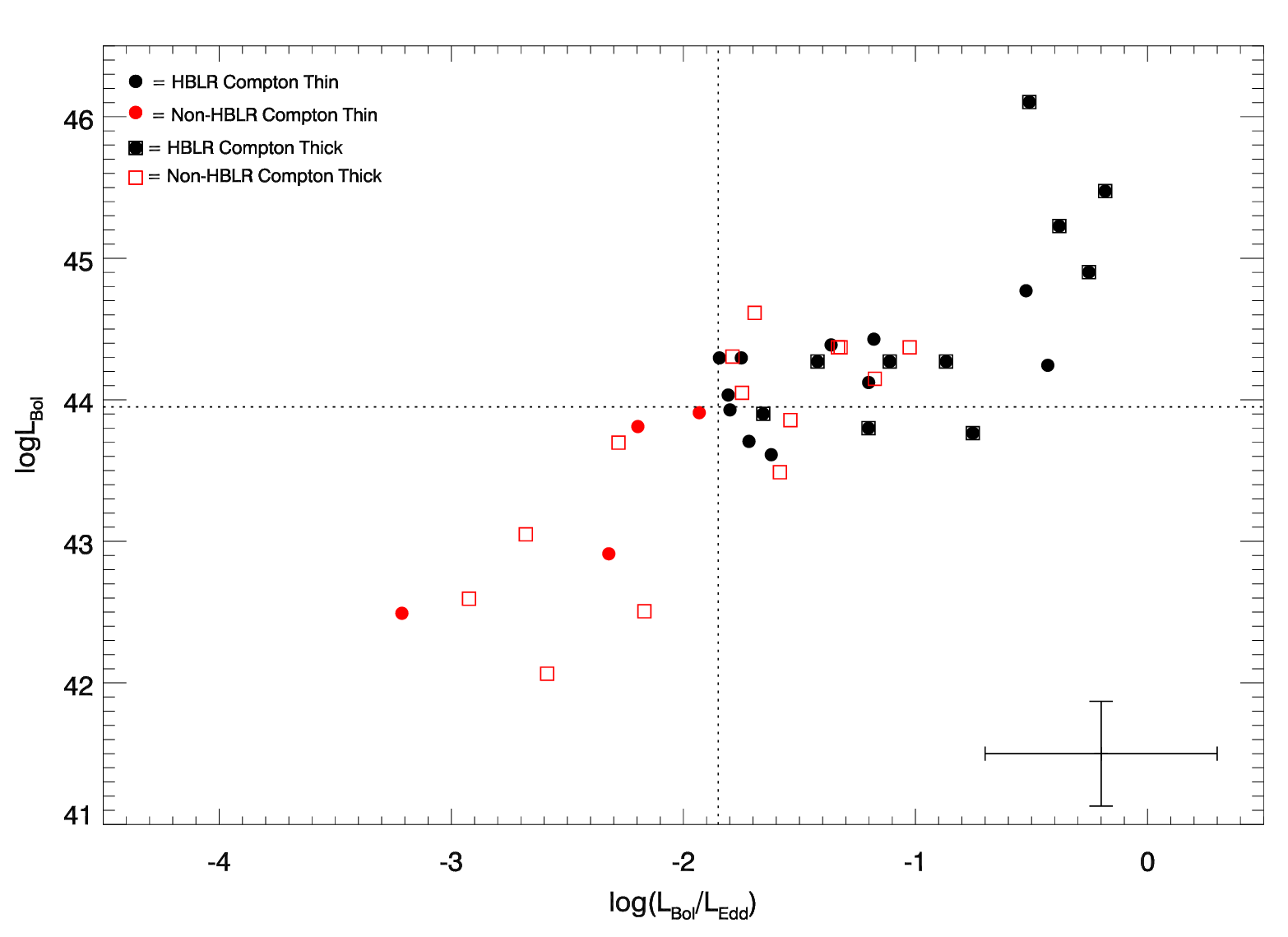}
\caption{\label{lumbol_mdot} The bolometric luminosity inferred from the 2-10 keV luminosity against the Eddington ratio for all the Compton-thin sources of our sample is plot in the top panel while in the bottom plot Compton-thick sources are introduced, using a correction factor of 70. The dotted lines represent the maximal separation values between the cumulative distribution of the two samples with respect to $L_{Bol}$ and $\lambda_{edd}$.}
\end{figure*}

We first considered only the Compton-thin sources of our sample. 
In the top panel of Fig.~\ref{lumbol_mdot}, we plot the bolometric luminosity against the Eddington ratio for all the Compton-thin sources 
of our sample. 
HBLR and non-HBLR sources are clearly separated both in luminosity and in accretion rate. 
To address the statistical significance of this separation (both in luminosity and accretion rate), we performed a two-sample 
Kolmogorov-Smirnov test on our data. These tests give probabilities of 0.5\% (luminosity) and 0.1\% (Eddington ratio) that the 
two classes are taken from the same parent population. 
These correspond to 2-sided Gaussian-equivalent significances of 2.8$\sigma$ and 3.3$\sigma$, respectively. 
The threshold Eddington ratio and luminosity that minimize the probability of chance separation, are 
$\log \lambda_{edd}=-1.9$ and $\log L_\mathrm{bol}=43.90$, respectively. 

We then repeated our analysis, by including also Compton-thick sources (for which inferred intrinsic 2-10 keV luminosity estimates are 
uncertain: see Sect.~\ref{xanalysis}). The inclusion of Compton-thick sources makes the HBLR versus non-HBLR separation 
less sharp (Fig.~\ref{lumbol_mdot}, bottom panel), and less significant, with probabilities of chance separation 
of 1.0\% and 1.8\%, for Eddington ratio and bolometric luminosity, respectively (Table~\ref{kstests}), corresponding to statistical 
significances of only 2.4$\sigma$ and 2.6$\sigma$. 
However, we note that no HBLR source falls below the Eddington ratio threshold. 

\subsection{Comparison with other Works and Methodologies}
Other authors investigated larger (but, in most cases inhomogeneous) samples to search for physical parameters responsible for the 
existence or absence of HBLRs in AGNs, and found often no clear indication for the existence of such a parameter \citep[e.g.][]{biangu07}. 
However, we think that this is due to the non-homogenous nature of the assembled samples, or to the ill-suited methodologies 
used to derive bolometric luminosities and/or BH masses, or to a combination of both issues. 

\begin{figure}
\centering
\includegraphics[width=5in]{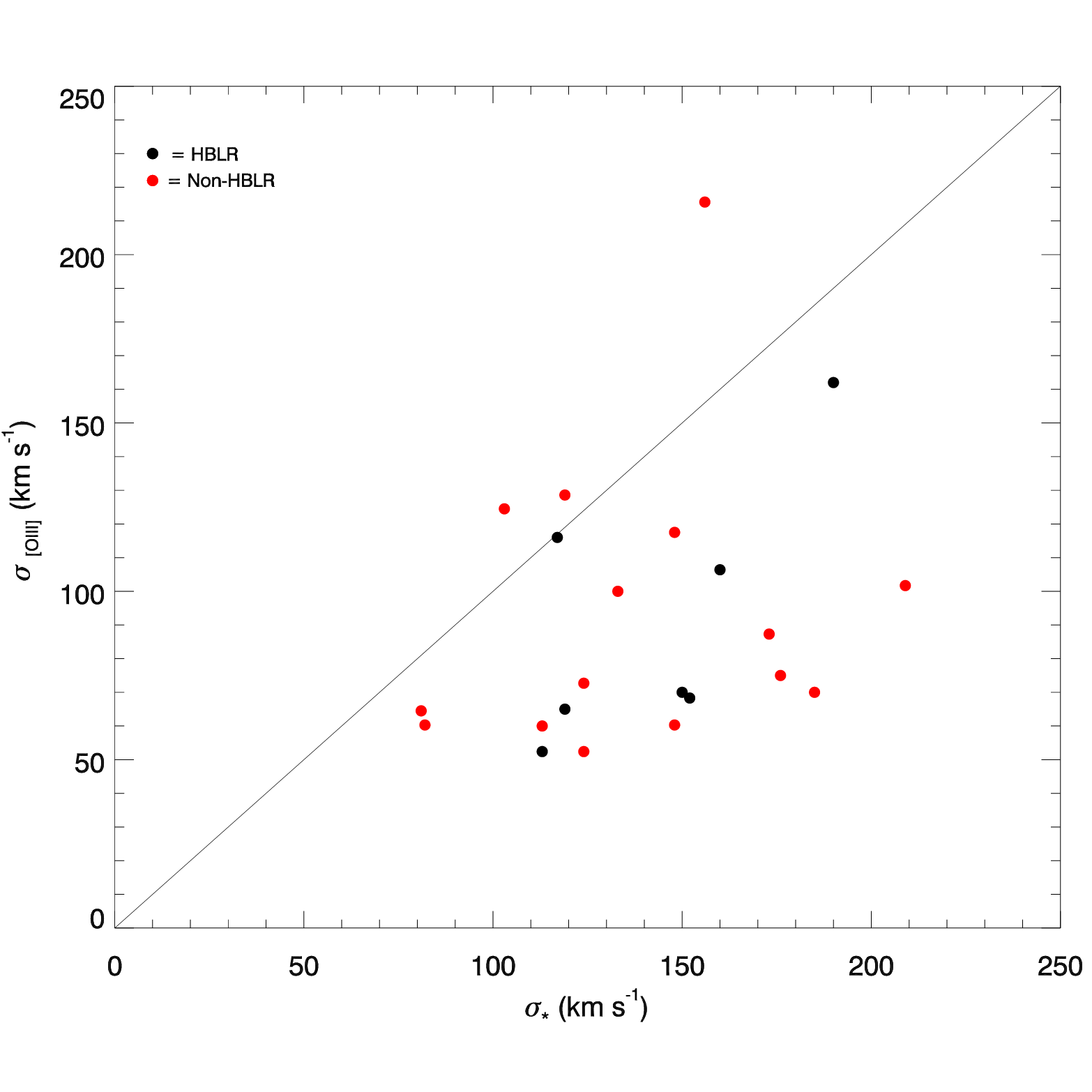}
\caption{\label{oiiiBH}Stellar velocity dispersions obtained from the Ca\textsc{ii} and Mg b absorption lines against those derived through the FHWM of the [O\textsc{iii}] emission line (on the Y axis). The two estimates are very different and not correlated. Only the BH masses estimated with the first method will be used in this work.}
\end{figure}

BH masses can be derived through a variety of different means (e.g. FWHM[OIII], $\sigma^\star$). Using all of these possible means, obviously 
allows for large samples to be assembled. However, building samples with black-hole mass determinations derived with different 
methods, introduces significant sources of uncertainties, which may wash out possible correlations among important physical 
parameters, like the source accretion rate or luminosity. 
For example, stellar velocity dispersion can in principle be inferred through the relation $\sigma_*=($FWHM$_{[OIII]}/2.35)\times 
1/1.34$ \citep{gh05}, by making use of [O\textsc{iii}] FWHM measurements, which are available for a large number of sources in the 
literaure. We tested the goodness of this estimator against the sources of our sample, for which direct stellar velocity dispersions are 
available. For all sources of our sample, we searched the literature for [O\textsc{iii}] FWHM measurements, and applied the \citet{gh05} 
relation to derive the stellar velocity dispersions. We then compared these estimates with the measurements of $\sigma_*$. 
Fig.~\ref{oiiiBH} shows the result of this comparison: clearly the two values differ significantly and do not appear to be 
linearly correlated with one another. We, therefore, use only BH estimates based on direct measures of the $\sigma_*$, in order to avoid further uncertainties due to the indirect derivation of this parameter.

\begin{figure}[t!]
\centering
\includegraphics[width=5in]{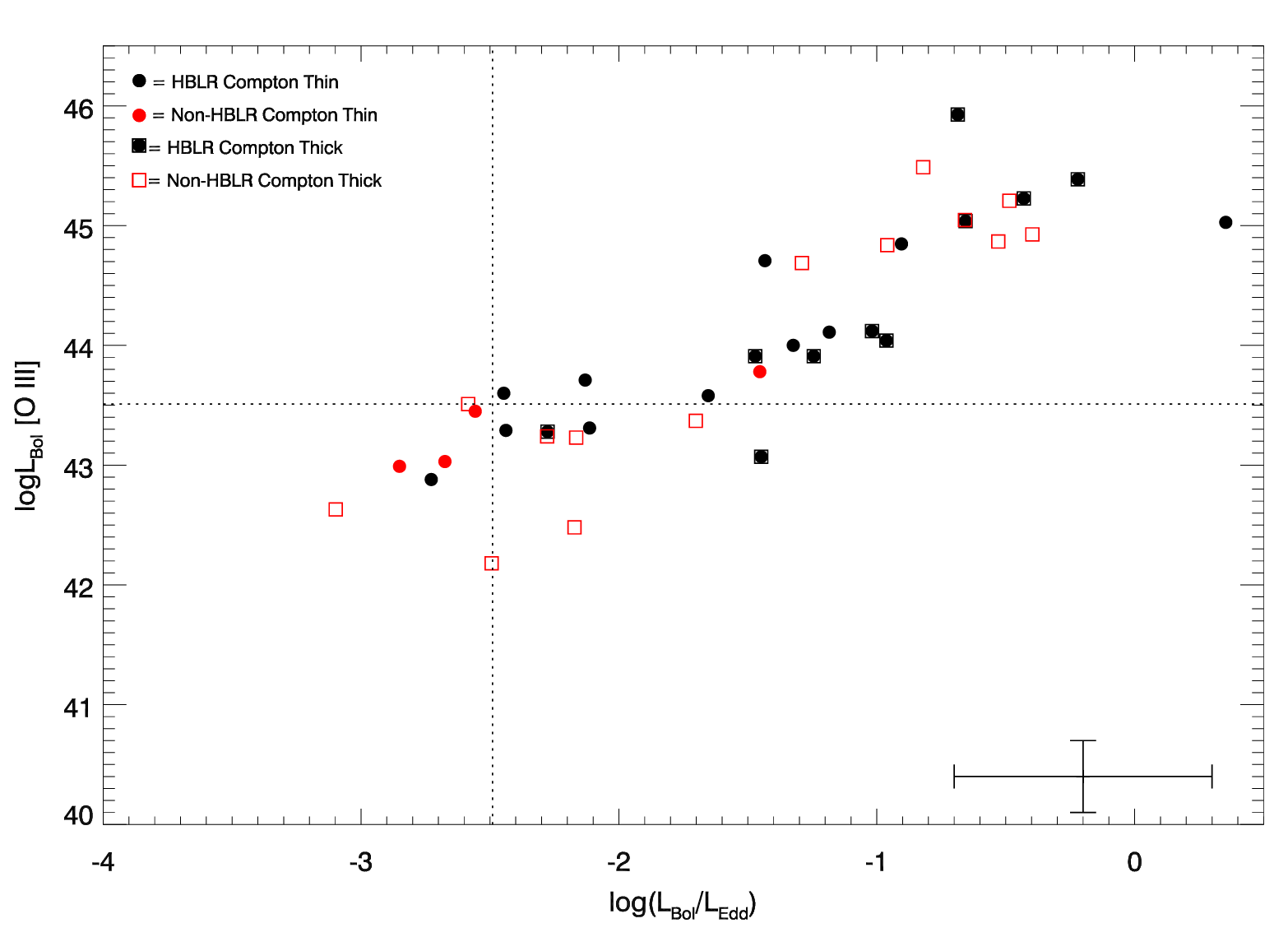}
\includegraphics[width=5in]{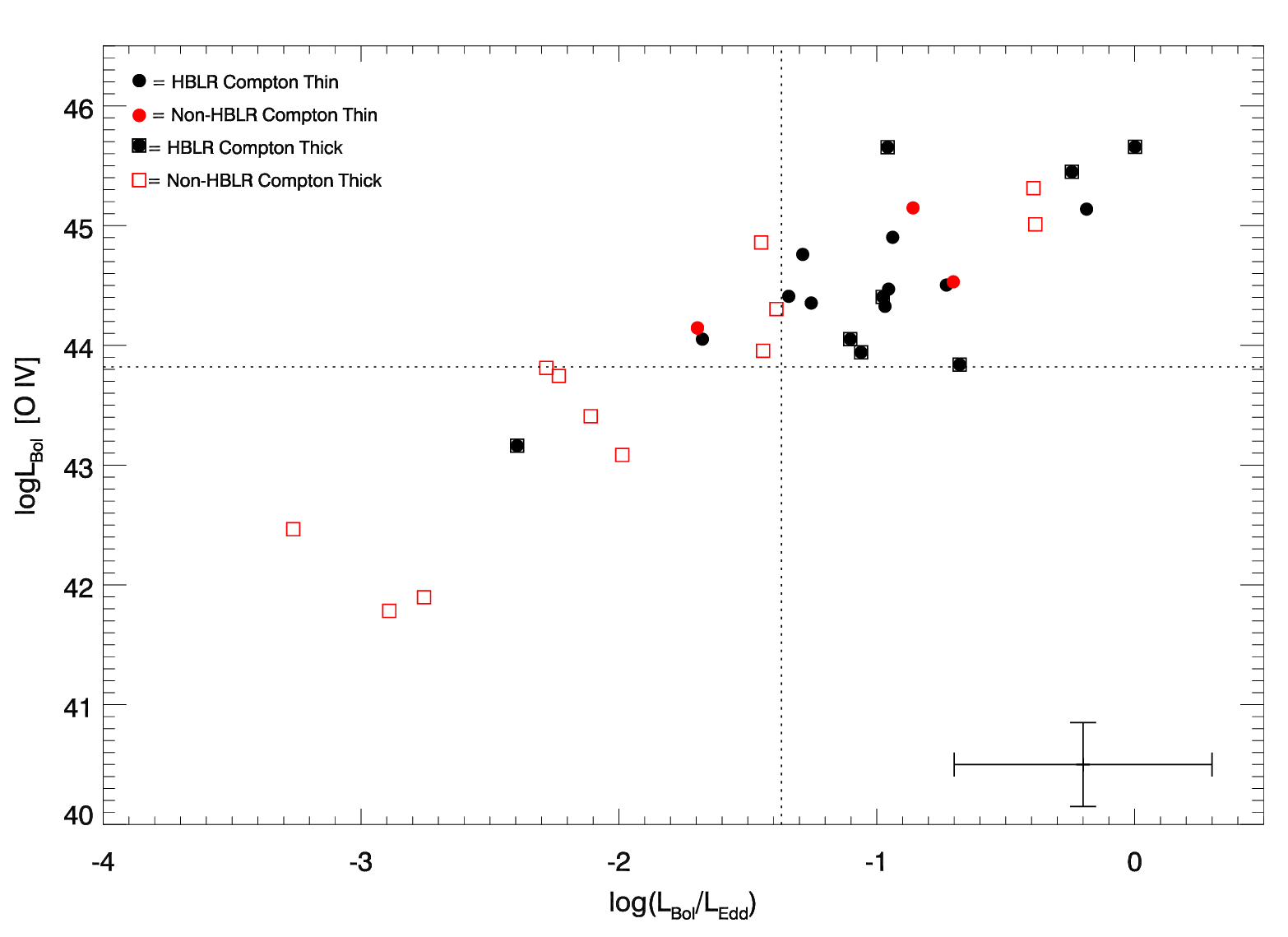}
\caption{\label{lumoiii_mdot} In the top panel bolometric luminosities derived from the [O\textsc{iii}] and Eddington ratios for all the 39 Compton-thin and Compton-thick sources of our sample are shown. In the bottom panel bolometric luminosities derived from [O\textsc{iv}] and Eddington ratios for 32 sources are shown. We used a dispersion value of $\sigma=0.3$ dex for uncertainties in the estimates of the [O\textsc{iii}]  bolometric luminosities, based on a recent study presented in \citet{risa11} where the flux of the [O\textsc{iii}] line is used as a reliable indicator of the bolometric emission of quasars. For [O\textsc{iv}] bolometric luminosities we used a dispersion value of $\sigma=0.35$ dex \citep{goual10}. The dotted lines represent the maximal separation values between the cumulative distribution of the two samples with respect to $L_{Bol}$ and $\lambda_{edd}$.}
\end{figure}

Another important source of potential error, is the proxy used to derive the nuclear bolometric luminosity of an obscured AGN. 
In our work, we use 2-10 keV luminosities, which are a direct probe of the intrinisic nuclear activity. 
However, such luminosities are available only for a limited number of Seyfert 2s. 
On the contrary [O\textsc{iii}] luminosities only echoes the intrinsic nuclear activity, but are available for a much larger number of 
obscured AGNs. 
To check the goodness of the [O\textsc{iii}] luminosity estimator, we compared our results with those obtained, for the sources of 
our sample, by using the the [O\textsc{iii}] luminosities and the bolometric luminosities derived from them, through the bolometric correction $C_{[OIII]}$ described in \citet{lamastra09}. The results are shown in 
Table~\ref{kstests} and Fig~\ref{lumoiii_mdot} (top panel): HBLR and non-HBLR sources are now mixed in both bolometric luminosities and 
Eddington ratios, with no significant separation between the two classes. 
The presence or absence of a clear separation, either in acrretion rates or bolometric luminosities, between Compton-thin 
HBLRs and non-HBLRs, when these quantities are inferred from 2-10 keV (probabilities of chance separation of 0.1\% and 0.5\%, 
for $\lambda_{Edd}$ and $L_{Bol}$, respectively) or [O\textsc{iii}] (probabilities of chance separation of 8.7\% and 38.5\%) observed 
luminosities,  confirms  \citep[e.g.][and references therein]{lamastra09, trou10} that the [O\textsc{iii}] luminosity is not a direct proxy 
for the bolometric luminosity of an AGN. \\
A large number of sources in our sample (32 out of 39) also have [O\textsc{iv}] measurements \citep{pere10, weav10}.  For a few sources we inferred [O\textsc{iv}] luminosities following the method presented in \citet{goual09}. Using the relation in \citet{goual10}, both bolometric luminosities and accretion rates can be inferred from the [O\textsc{iv}] luminosities. The results are shown in the bottom panel of Fig. \ref{lumoiii_mdot} and Table \ref{kstests}. The two classes show a more significant separation in accretion rate than in bolometric luminosity (probability of chance separation of 1.1\% and 10.4\%, respectively\footnote{A K-S test for the subsample of only Compton-thin sources cannot be performed, since the minimum number of data values for a statistical significance is 4 \citep[pag. 120-121]{ste70}}). Both HBLR and non-HBLR Compton thick sources are included in the sample and the separation between the two different populations in accretion rate resembles the one inferred from our X-ray analysis, suggesting that the efforts to use standard 
extinction correction for [O\textsc{iii}] are not always reliable \citep[e.g.][]{goual09, lamassa10}.

\section{Discussion}
Since the seminal results by \citet{miller83} on the archetypical Seyfert 2 galaxy, NGC1068, many efforts have been spent to understand 
whether the Unification Model is valid for all Seyfert galaxies, or the presence/absence of the BLR may be also linked to intrinsic properties 
of different classes of sources. 
Intrinsic differences were strongly advocated by \citet{tran01} and \citet{tran03}, who found that HBLR Seyfert 2s are normally associated 
with typical obscured Seyfert 1 nuclei, while non-HBLR Seyfert 2 galaxies host, on average, significantly weaker muclei, likely incapable 
of generating classical BLRs.

This evidence supports theoretical models that link the presence or absence of BLRs in Seyferts, to intrinsic nuclear properties. 
In a model proposed by \citet{nic00}, the presence of broad emission lines is intrinsically connected with disk instabilities occuring 
in proximity of a transition radius at which the accretion disk changes from gas-pressure dominated to radiation-pressure dominated. 
The transition radius is function of the accretion rate, and becomes smaller than the innermost stable orbit for accretion rates (and therefore 
luminosities) lower than a threshold that depends weakly on the BH mass. Weak AGN should, therefore, lack the BLR. 

In this work we try to test this model in the least biased observational way, by collecting the 'cleanest' possible sample of Seyfert 2 
galaxies with the best spectropolarimetric data available at the moment, and robust estimates of their BH masses performed 
homogeneously on the basis of the observed stellar velocity dispersion. 
When only Compton-thin sources are considered, the estimate on their accretion rate is done directly on the observed nuclear 
X-ray emission, and the modest absorption is likely not able to affect severely the detection of the hidden BLR, if present. 
We find that the separation between HBLR and non-HBLR Compton-thin sources is highly significant both in accretion rate and 
luminosity. In particular, no HBLR is found below the threshold Eddington rate $\log \lambda_{edd}=-1.9$, and no 
non-HBLR above the same limit. 
Even when Compton-thick sources are included, no HBLR is still found at accretion rates lower than 
the above threshold. 
 This threshold accretion rate is in good agreement with the value presented in a recent work \citep{trum11}. 

 This result supports the theoretical expectations of the model proposed by \citet{nic00}, albeit with a slightly 
higher value of the threshold accretion rate ($\log \lambda_{edd} \sim -2.5$ in \citet{nic00}). 
We stress here that the model proposed by \citet{nic00} applies only to radiatively efficient AGNs, accreting through a Shakura-Sunyaev 
disk \citep[SS-disk  hereinafter,][]{ss73}. 
Broad Emission lines are known to exist in some objects \citep[mostly LINERs or transition objects, e.g.][]{eliho09} accreting 
at rates as low as $\log \lambda_{edd} \sim -6$ \citep[e.g. M~81,][]{ho08}. These objects are highly radiatevely inefficient, and therefore 
most likely do not host a classic SS-disk.

If BLRs do not exist in weakly accreting AGN, one would expect the existence of unabsorbed, genuine Seyfert 2 galaxies. 
Such objects do indeed exist, and the best examples (where the lack of optical broad lines and of X-ray obscuration are 
unambiguosly assessed in simultaneous high S/N observations) have Eddington rates lower than the threshold estimated in this 
paper: NGC~3147 \citep[$\log \lambda_{edd} \simeq-4$:][]{bianchi08a},
Q2131427 \citep[$\log\lambda_{edd}\simeq-2.6$:][]{panessa09}, NGC~3660 
\citep[$\log \lambda_{edd}\simeq-2$: Bianchi et al., in preparation;][]{brna08}.

\noindent
It is interesting to note that among all the 156 X-ray unobscured ($N_H<2\times10^{22}$ cm$^{-2}$) AGN observed with 
XMM-\textit{Newton} \citep[CAIXA:][]{bianchi09}, only 6 have an Eddington rate lower than the threshold found in this 
work\footnote{An updated catalogue was used for this search.}. 
Four of these 6 sources were previously classified as unabsorbed Seyfert 2s \citep{pb02,panessa09,brna08} but see also 
\citet[e.g.][and references therein]{shi10}, who recently claimed the presence of very broad and weak broad emission lines in two 
of these objects). 
The remaining two objects are PG~1011-040 and NGC~7213, and both show the presence of BLR in their optical spectra. 
Both objects are peculiar. PG~1011-040 
is only slightly below the accretion rate threshold of $\log \lambda_{edd}\simeq-1.9$ that we find here and, most 
importanly, is severely under-luminous in X-rays with respect to its multiwavelength luminosity, so that the accretion rate derived 
from its 2-10 keV luminosity by simply applying the \citet{mar04} correction, is likely severely under-estimated \citep{gal01,vf07}.
NGC~7213 is the only bright Seyfert 1 galaxy known to unambiguously lack reprocessing features 
from Compton-thick distant material \citep{bianchi03b,bianchi04,lob10}.

The inclusion of Compton-thick sources in our sample, makes the HBLR versus non-HBLR separation, in terms of either Eddington ratios 
or bolometric luminosities, less significant. Several non-HBLR Compton-thick sources lie now in the HBLR $L_{Bol}$ -- $\lambda_{Edd}$ 
plane, i.e. at accretion rate higher than the threshold we find for Compton-thin sources. 
This mixing could be, at least partially, due to the difficulties in assessing the intrinsic 2-10 keV luminosity of 
Compton-thick sources, because of the unknown geometry of the absorbers and reflectors 
\footnote{The same uncertainties are probably also responsible for the lack of HBLR and non-HBLR separation when indirect estimators of the 
nuclear bolometric luminosity (e.g. $[O\textsc{iii}]$) are used.}
 
However, such uncertainties should in principle affect equally over- and under-estimates of the luminosities.
Instead, while no Compton-thick HBLR is still found below the Eddington rate threshold, a significant fraction (64\%) of Compton-thick 
non-HBLR has accretion rates higher than this limit. 
Moreover, uncertainties in the exact geometry of absorbers and reflectors, could only be responsible for modest (factor of 2-3) 
over-estimates of the intrinsic bolometric luminosity and so of the Eddington accretion rate. This could perhaps explain the presence of 
few non-HBLR closely above the Eddington ratio threshold, but can hardly explain order-of-magnitude over-estimations of the 
bolometric luminosity, as inferred from the 2-10 keV spectra. 

If the scenario proposed by \citet{nic00} is correct, non-HBLR Compton-thick sources with Eddington accretion rate estimates 
much above the threshold, are likely to be peculiar objects, where BLR line emission cannot be seen, not even in polarized light. 
These source should possess a BLR, but something prevents us from observing it. Different, alternative scenarios, to the intrinsic lack 
of a BLR, have already been proposed in the past \citep{lumsden01, guhuang02, mama02, tran03, moran07}.
This could be either because of the lack of an unobscured population of hot electrons scattering the line emission along our line 
of sight, or because the orientation of these sources is such that our line of sight intercept larger portions of absorbing gas, and this 
covers both the nuclear source and at least part of the line-reprocessing electron region \citep[e.g.][and references therein]{shu07, wu11}. 

\section{Conclusions}
The main results of this paper can be summarized as follows:\\
\begin{itemize}
\item{we presented evidence suggesting that accretion rate is the main ingredient which drives the presence of HBLRs in Seyfert 2 galaxies. 
By selecting a sample of 39 type 2 AGN with good quality spectropolarimetric and X-ray data, and for which we derived homogenous 
estimates of the mass of the central supermassive BH, 
we found a clear separation between Compton-thin HBLR and non-HBLR sources, both in luminosity ($\log L_\mathrm{bol}=43.90$) and 
in accretion rate ($\log \lambda_{edd}=-1.9$). A statistically similar separation is seen when bolometric luminosities are derived from infrared [O\textsc{iv}] emission lines.\\
Our results agree with the ones discussed in \citet{nmm03} but have higher statistical significance (due to the larger sample we use here, compared to \citet{nmm03}), and probably more robust due to the more accurate estimates of the BH masses that we derive here;}

\item{the inclusion of Compton-thick sources, washes out the separation between HBLR and non-HBLR, but still no HBLR source falls below 
the Eddington ratio threshold. 
We propose that the presence of a significant fraction (64\%) of Compton-thick non-HBLRs at accretion rates higher than the 
threshold found for Compton-thin sources, is not due to the lack of BLRs, but to heavy line-of-sight absorption preventing us 
from observing not only the direct line emission, but also their polarized light scattered off an, at least partly, obscured 
population of hot electrons. }
\end{itemize}
\acknowledgements AM, SB and GM acknowledge financial support from ASI under grant I/088/06/0. 
FN acknowledge financial support from ASI grants AAE and ADAE, as well as NASA grant NNG04GD49G. 
We would like to thank M. Elvis, for useful advices and discussions. We thank the anonymous referee for his/her important comments that greatly improved this work.\\
\newpage
\bibliographystyle{apj}
\bibliography{sbs} 

\end{document}